\documentclass[conference]{IEEEtran}
\IEEEoverridecommandlockouts

\usepackage[utf8]{inputenc}
\usepackage{lipsum}
\usepackage{tabularx,booktabs}
\usepackage{rotating}
\usepackage{dingbat}
\usepackage{makecell}
\usepackage{soul}
\usepackage[symbol]{footmisc}
\usepackage[table, x11names]{xcolor}
\usepackage{tablefootnote}
\usepackage{todonotes}

\def\BibTeX{{\rm B\kern-.05em{\sc i\kern-.025em b}\kern-.08em
    T\kern-.1667em\lower.7ex\hbox{E}\kern-.125emX}}
\graphicspath{ {./figures/} }

\begin{document}

\title{Online Malware Classification with System-Wide \\ System Calls in Cloud IaaS}

%\iffalse
\author{\IEEEauthorblockN{Phillip Brown, Austin Brown and Maanak Gupta}
\IEEEauthorblockA{\textit{Department of Computer Science} \\
\textit{Tennessee Tech University} \\
Cookeville, TN USA \\
pabrown43@tntech.edu, ambrown51@tntech.edu, mgupta@tntech.edu}
\and
\IEEEauthorblockN{Mahmoud Abdelsalam}
\IEEEauthorblockA{\textit{Department of Computer Science} \\
\textit{North Carolina A\&T State University}\\
Greensboro, NC, USA \\
mabdelsalam1@ncat.edu}}
%\fi
\maketitle

\begin{abstract}
Accurately classifying malware in an environment allows the creation of better response and remediation strategies by cyber analysts. 
However, classifying malware in a live environment is a difficult task due to the large number of system data sources. 
Collecting statistics from these separate sources and processing them together in a form that can be used by a machine learning model is difficult. 
Fortunately, all of these resources are mediated by the operating system's kernel. 
User programs, malware included, interacts with system resources by making requests to the kernel with \textit{system calls}. 
Collecting these system calls provide insight to the interaction with many system resources in a single location. 
Feeding these system calls into a performant model such as a random forest allows fast, accurate classification in certain situations. 
In this paper, we evaluate the feasibility of using system call sequences for online malware classification in both low-activity and heavy-use Cloud IaaS. 
We collect system calls as they are received by the kernel and take n-gram sequences of calls to use as features for tree-based machine learning models. 
We discuss the performance of the models on baseline systems with no extra running services and systems under heavy load and the performance gap between them.
\end{abstract}


\section{Introduction}

The rapid infrastructure churn of modern cloud systems requires a fast, scalable malware classification system that provides actionable intelligence that can be used for rapid remediation. In this work, we attempt to design a malware classification system that is able to run on features collected in real time from a running, online Linux system. Our primary motivation is collection performance: feature collection must not interfere with the processes on the system that are required for business value. Our chosen primary features, system calls, are more difficult to collect than simpler summary statistics like performance metrics or network flows, but their granularity and breadth of information are wonderful aids when classifying malware \cite{syscalls-usefulness}. 

Malware analysis is broadly divided into two disciplines: static and dynamic. Static malware analysis deals with malware at rest. Static analysis inspects malware files without executing them. This makes static analysis a safe method of analysis because malware cannot damage an analysis machine or network. %Static features included metrics such as file section names and sizes, API call counts, byte sequences, and control flow. These features are safe and easy to collect, i
Modern malware creation techniques can often evade static analysis by hiding parts of the executable or disguising control flow. 
%Binary packing compresses and encrypts the machine instructions in an executable. This may be used legitimately to reduce the size of an executable for distribution, but is more often used to hide the code of an executable from malware analysis. 
The ease of disguising malicious code from static analysis is the greatest weakness of static analysis.

Dynamic analysis studies malware during and after execution. Unlike static analysis, the malware is actively executed and its effects on the system are studied. Much more information about a piece of malware can be collected during dynamic analysis. Since executable sections must be available for the malware to run, dynamic analysis defeats packing strategies that would evade static analysis. Further, network, memory, and file activity on the system can be observed to investigate what parts of the system the malware is affecting.

Malware is often destructive so care must be taken to create a network where damage that may be caused by the malware is reversible and contained to a well-monitored and segmented part of the network. Malware may attempt to detect if it is running in a sandbox or being monitored and alter its behavior to avoid analysis. Careful design of the execution environment can reduce the malware's ability to detect that it is being inspected. \textit{Online} dynamic analysis is a specific dynamic technique where the malware is not run in a simulated environment but rather a real, internet connected system. This allows malware to connect to internet resources if those are required for its operation. While this gives the best view into the behavior of the malware, it is the most dangerous method of analysis, as giving malware access to the internet potentially allows it to spread and infect vulnerable systems. We have specifically designed our experiments and environments to leverage online analysis to replicate real enterprise installations that may be targeted by malware authors.

\begin{table*}[h]
\label{Related Works}
\centering
\caption{Related Works Compared}
\begin{tabularx}{.70\textwidth}{|c||c|c||c|c||c|c||c||c|c|c|}\hline
     Reference & \multicolumn{2}{c||}{Analysis} & \multicolumn{2}{c||}{Features} & \multicolumn{2}{c||}{ML Technique} & Domain & \multicolumn{3}{c|}{Platform}\\\hline
     & \begin{turn}{90}Classification\end{turn} & \begin{turn}{90}Detection\end{turn} & \begin{turn}{90}System Calls\end{turn} & \begin{turn}{90}Other\end{turn} & \begin{turn}{90}Traditional\end{turn} & \begin{turn}{90}Deep Learn.\end{turn}  & \begin{turn}{90}Online\end{turn} & \begin{turn}{90}Linux\end{turn} & \begin{turn}{90}Android\end{turn} & \begin{turn}{90}Windows\end{turn} \\\hline
     Canzanese \textit{et al.} \cite{canzanese-automatic-online-classification} & \checkmark & & \checkmark & & \checkmark & & & & & \checkmark\\\hline
     Babenko and Kirillov \cite{babenko-stat-methods-and-extended-syscalls} & \checkmark & & \checkmark & & & \checkmark & & & & \checkmark \\\hline
     Canfora \textit{et al} \cite{canfora-android-syscall-sequences} & \checkmark & & \checkmark & & \checkmark & & \checkmark & & \checkmark & \\\hline
     Alsulami and Mancoridis \cite{alsulami-prefetch-cnn-classification} & \checkmark & & & \checkmark & & \checkmark & & & & \checkmark \\\hline
     Das \textit{et al} \cite{das-fpga-semantics-detection} & & \checkmark & \checkmark & & & \checkmark & & \checkmark & & \\\hline
     Chandramohan \textit{et al} \cite{chandramohan-bofm-detection} & & \checkmark & \checkmark & & \checkmark & & & & & \checkmark \\\hline
     Dawson \textit{et al} \cite{dawson-phase-space-detection-syscalls} & & \checkmark & \checkmark & & \checkmark & & & \checkmark & & \\\hline
     Azmandian \textit{et al} \cite{azmandian-vmm-ids} & & \checkmark & & \checkmark & & & & & & \checkmark \\\hline
     
     \textbf{Our approach} & \checkmark & & \checkmark & & \checkmark & & \checkmark & \checkmark & & \\\hline
\end{tabularx}
\footnotetext{Experiments performed on a physical Android phone with network access}
\end{table*}

\section{Motivation}
We chose to perform malware classification over detection because it gives greater insight into the threat and scale of possible damages for an infection. Simple spam tools or adware may not dictate the same measure of response that a crypto-locker or remote access Trojan would. %It may be tolerable on a network for a short period while careful mitigation is performed to not disturb running business processes, while the latter malware likely necessitates emergency measures that may disturb intended functionality for a short while in order to prevent spread and damage  rest of the network.
Our specific dynamic analysis implementation is intended to detect a compromised system by inspecting system-wide features. %Single-process analysis is likely much more effective at direct classification because there is no other activity to consider. 
While one process may be easier to classify, doing classification for every process on a system would quickly become impossible. Therefore, we observe features from the entire system at once. This has the additional benefit of being able to detect the compromise of existing benign processes on the system. "File-less" malware is malicious code that is somehow injected into the process of an already running process. This avoids malware detection if the detection is observing only new processes or has already flagged the compromised process as benign.

We chose to target online Linux systems because of the current and likely future popularity for the GNU/Linux-based operating systems in cloud computing. Cloud environments are increasingly popular for enterprise use because of their scalability, price efficiency, and availability. As their popularity increases, these systems are increasingly being targeted by malware authors seeking to compromise them. %Finding fast and reactive ways to detect compromise early and form a remediation plan would save money and time for cloud operators and clients.

Our choice of model and features were also driven by practical considerations. While modern neural models have shown great promise for security work, their high resource requirements during training and the high number of samples required to effectively train them makes them difficult to implement without considerable infrastructure investment. Traditional machine learning models require less training data and can be effectively trained on consumer-grade CPU hardware, and their performance has been proven to be acceptable for security tasks \cite{classifiers-compared, classifier-performance-eval}.

The main contributions of this work are:
\begin{itemize}
    \item We evaluate the feasibility of using traditional machine learning models for classifying malware from behavior in a custom whole-system system-call sequence data set from simulated cloud IaaS systems.
    \item We show this approach is effective for low-activity cloud systems, but is less effective when systems are under heavy load and generating many system calls.
\end{itemize}

\section{Related Work}
Malware detection and classification are wide and active disciplines. To narrow the field of peers, we will compare our work to only those other works that are attempting \textit{dynamic} analysis. While there is a plethora of static malware analysis research, static and dynamic analysis methods are distinct enough that static analysis is not comparable to our work. Static analysis analyzes the attributes of malware files while dynamic analysis is focused on observing the behavior of a malware during its execution. We compare our work to only those other works that are also observing malware behavior. We would be particularly interested in comparing our work to other work where the malware is not sand-boxed and is able to run as it would "in the wild" on compromised systems, but such research is understandably difficult to find due to the dangers of running malware on live systems.
%and the (completely reasonable) reluctance of most institutions to allow such research on their systems. 
%Our access to an open cloud platform allowing online malware research is a primary advantage of our research because it allows us to collect behavioral artifacts from malware that are exactly the same as if the malware had compromised a real-world system.

%While there exists current dynamic feature research on Linux malware, dynamic feature classification, and online malware analysis, it is difficult to find research that combines all three as our research does. 
%There exists much more work on malware \textit{detection} than exists work on malware \textit{classification}, perhaps due to the difficulty in acquiring data sets with usable class distributions which we discuss in section \ref{sample-processing}.
There is a large body of work on malware detection with various sources of dynamic features collected at run-time. We can make useful comparisons here to data collection and processing because these are largely similar between detection and classification tasks. %Differences between the approaches arise in the machine learning design and implementation phases, however, but these related works were still instructive for how we performed our data processing. 
Reference \cite{dawson-phase-space-detection-syscalls} uses VM introspection tools to collect system calls from a VM running below the context of the monitor system. 
%The authors chose to use the \textit{strace} utility, which we found to be too performance intensive to reasonably use on a live system because it invokes a process interrupt on every system call. 
Reference \cite{chandramohan-bofm-detection} provides an insightful method to process raw system calls into a format better suited for machine learning. 
%Raw system call traces are usually a stream of text entries that describe every interaction a process has with the system kernel. 
%The BOFM method proposed in this work extracts important statistical features from system call streams and generates a feature space that is complete but small enough to reasonably process without exotic and expensive hardware. 

\subsection{System Call Collection}

%References \cite{canzanese-automatic-online-classification} and \cite{das-fpga-semantics-detection} focus on Linux malware classification using system calls. 
Reference \cite{das-fpga-semantics-detection} is perhaps the most similar in concept to our own, also collecting Linux system calls. 
This work, however, is only concerned with single isolated systems and does not classify malware, only detects it.  
%Also, they offload the machine-learning detection task to a FPGA chip separate from the system that the malware infected. While an interesting technique with several advantages related to performance and security, it is not practical for cloud-native systems.
Reference \cite{canzanese-automatic-online-classification} collects system calls for malware classification, but does so on hosts running the Windows operating system that are not connected to the internet. 
%Importantly, the authors call their approach "online," but they are referring to the machine-learning definition of online: their system continuously learns and adapts to new samples. This differs from our definition of online where the system that malware samples are run on has unrestricted access to the internet and behaves just as a real-world system would. However, this work contains valuable insight on collecting system behavior statistics and using them in ML models, so we include it here.

Several malware detection works focus on the Android mobile operating system. %As of 2021\footnote{https://gs.statcounter.com/os-market-share/mobile/worldwide}, Android is the most used mobile OS by a wide margin and its usage numbers even challenge existing desktop numbers. %Based on the Linux kernel, Android has become a successful operating system and has drawn attention from both malware authors and security researchers. 
While Android applications are packaged and managed differently than mainline GNU/Linux distributions, the similarities in the platforms make Android security research valuable contributors to our own work. Reference \cite{canfora-android-syscall-sequences} uses system call sequences collected from Android program execution to detect malware. Their approach uses n-grams of system calls and the counts of those n-grams as features for a support vector machine. Similar to our work, the authors in this work wanted a realistic environment for malware to execute. To this end, they executed their malware and benign data sets on a physical Android phone instead of using Android emulators.

\section{Data Collection Methodology}

\subsection{Experiment Environment}\label{exp-env}

A primary goal of this research is to design malware classification that works in real environments. All of the malware experiments were carried out on real Linux cloud servers with unrestricted internet access. The experiments were hosted on an OpenStack\footnote{www.openstack.org} instance graciously provided by the University of Texas at San Antonio\footnote{www.utsa.edu}. OpenStack provides a free and powerful cloud infrastructure platform like those that are popular in industry and as targets for malware authors. This OpenStack platform emulates production cloud environments with full Linux systems. This provides some important advantages for malware analysis over safer methods like sand-boxing or emulation; the systems used in this research are just like those in the wild that malware will be written to infect. These systems likely pass most network and sandbox-detection evasion methods in malware so that the potentially hidden behavior of the executable can be captured.

\begin{figure}[!t]
%\centering
    \includegraphics[width=.5\textwidth]{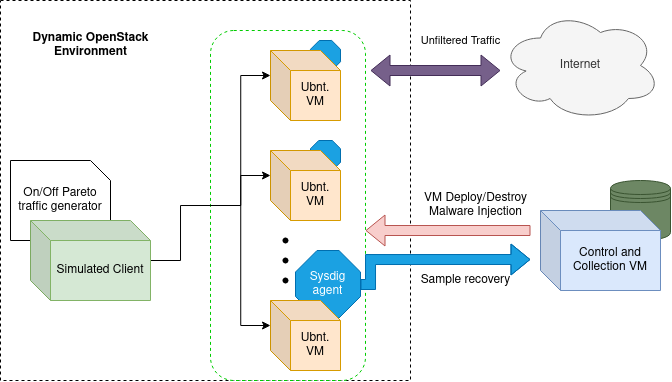}
    \caption{The OpenStack Environment where the malware was examined}
    \label{testbed}
\end{figure}
Malware behavior collection experiments were performed in two stages across the same environment. The environment, shown in Figure \ref{testbed}, is a number of target virtual machines (VMs) running fully-patched Ubuntu 18.04 on OpenStack, each with its own network connection and public IP address. %An additional VM acts as a controller node, orchestrating the creation and destruction of targets, the implanting of malware, the traffic generator, and the activity monitor. %Each target machine has identical hardware: 2 CPU cores, 4GB of RAM, and 40GB of disk space. The targets are spawned from an identical, fully patched Ubuntu 18.04 server image.
To collect malware behavioral data, each target machine was run for ten minutes. For the first five minutes, the machine ran unaltered. After approximately five minutes (some randomness was used to prevent malware from always executing at the same time in the traces), the controller node copies in and executes a malware executable. At ten minutes the target is destroyed and replaced by a new target and the experiment continues. In the first stage of experiments, no software other than what is required for the Ubuntu 18.04 image was running. In the second stage, the server was running an Apache web server hosting a WordPress site while simulated traffic was generated from the controller node. 
%The baseline first stage served to establish that malware behavior could be distinguished with system-wide monitoring in a relatively 'quiet' system with no other major processes running. The second stage was intended to determine if malware behavior could be detected among the noise of a system generating very many system calls because of high utilization.

%To collect system calls system-wide, a performant collection method was required. User-level system call collectors like the \verb|strace| utility must interrupt a process on every system call to intercept and record the call, making its performance impact considerable and inappropriate for our use. 
To efficiently collect the system calls from the live systems, we needed a kernel-level solution that did not interrupt process execution. The utility \verb|sysdig|\footnote{https://github.com/draios/sysdig} provides this functionality. System call collection is done in kernel at call time and is completely transparent to the calling process. Performance overhead is low enough to be acceptable in our experiments here, with no measured impact to application response.

During test execution on the targets, \verb|sysdig| was run to collect all system calls sent to the kernel from all processes. The output of each system call record was retrieved from the target machine, labeled with the hash of the malware executable used, and stored for later use. Results were stored as-is in the proprietary \verb|sysdig| binary output file because of its compact size and completeness. 
%While \verb|sysdig| collects many metrics in addition to system call names, we only use the \verb|syscall.type| field that records the system call name to produce the call sequences for this research.
\begin{table}[!t]
\caption{All classes from the 4,180 final samples}
\begin{center}
\begin{tabular}{l r r}
\toprule
Class       &   Count   &   Used        \\
trojan      &   2299    &   \checkmark  \\
virus       &   616     &   \checkmark  \\
backdoor    &   382     &   \checkmark  \\
rootkit     &   253     &   \checkmark  \\
miner       &   226     &   \checkmark  \\
grayware    &   142     &   \checkmark  \\
worm        &   142     &   \checkmark  \\
none        &   87      \\
ransomware  &   21      \\
downloader  &   7       \\
bot         &   3       \\
hoax        &   2       \\
\bottomrule
\end{tabular}
\label{virus-classes}
\end{center}
\vspace{-4mm}
\end{table}

\subsection{Executable Sample Selection and Curation} \label{sample-processing}
A major share of the work behind this research was collecting and curating a suitable malware set that provided good samples that ran in our selected environment and had enough class diversity to perform useful classification. Initially, the academic dataset from VirusTotal\footnote{https://www.virustotal.com/} was explored, but it was found to be poorly balanced. While the samples in the VirusTotal dataset were appropriate for our systems and were functional, the majority of them ($>$80\%) were Mirai or Mirai variants. %There were not enough samples from other classes to provide a good basis for training classification. 
Additional samples from MalShare and VirusShare\footnote{https://www.malshare.com/ and https://virusshare.com/} were added to increase the numbers of other classes. After removing duplicates and selecting only x86-64 binaries, from more than 10,000 total samples we selected 4,180 samples that were functional. These classes are shown in Table \ref{virus-classes} and those used in our experiments are check-marked. These classes were sufficient as a base for our final class selection described in Section \ref{machine-learning}.

\begin{figure}[!t]
    \includegraphics[keepaspectratio,width=\columnwidth]{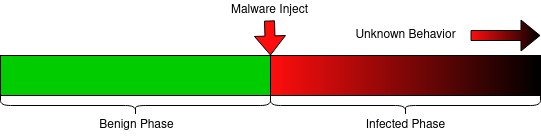}
    \caption{Behavior over time of the experiments}
    \vspace{-2mm}
\label{malware-over-time}
\end{figure}

\section{Machine Learning Classification}\label{machine-learning}
\subsection{Classes}\label{classes}
Our malware samples were not labeled by the providers. 
To determine a reasonable classification, we examined the combined output of the various engines behind VirusTotal with AVClass\footnote{https://github.com/malicialab/avclass} \cite{avclass-acsac}\cite{avclass-raid}, a tool written to determine family and class labels from VirusTotal scan results.

VirusTotal uses several engines and scanners to determine what a malware file is. 
At the time of writing, VirusTotal claimed over 70 anti-virus scanners and services.%\footnote{https://support.virustotal.com/hc/en-us/articles/115002126889-How-it-works}.
VirusTotal has an available API where an executable can be uploaded and VirusTotal will distribute the sample to all of those engines and return a report containing reports from each engine. 
These reports are not standardized; each AV vendor engine has their own reporting syntax and semantics. 
This creates a difficulty for determining accurate class labels for a piece of malware since it may be reported as different classes by different vendors, and each vendor may use a different term or label for the same class. 
While this makes determining proper labels more difficult, it is not impossible. 
Fortunately, AVClass can parse these VirusTotal API results and determine proper class labels.

Semantics have a large influence on malware classification. 
The classes in Table \ref{virus-classes} are labels created to describe malware based on some behavior that it exhibits. 
Some classes may have overlap in the behavior that they describe. 
In our data, we find the classes \verb|trojan|, \verb|virus|, and \verb|backdoor|. 
%From an analyst's perspective, these three classes are likely to describe malware that has similar behavior - namely infiltrating a system and providing some access to an external actor. 
Since the classification labels obtained from AVClass are sourced by consensus from many sources, the class labels are particularly likely to differ among the various engines based on the perspectives of those that wrote them. 
The \verb|trojan| class in particular was found by the authors of \cite{avclass-raid} to have often been used as a 'catch-all' class when a better classification could not be made. 
This has some impact on our classification work. 
%Overlap between classes can impact model accuracy without providing useful information for remediation or response.

The classes in Figure \ref{virus-classes} are the total set present in all the malware selected. 
Obviously, the low-numbered classes are not usable for training a machine learning model, so they are dropped along with the 'none' class samples that AVClass was not able to determine a class for. 
This left the seven classes \verb|trojan|, \verb|virus|, \verb|backdoor|, \verb|rootkit|, \verb|miner|, \verb|grayware|, and \verb|worm|. 
We did not pursue family attribution in this work, as we found that AVClass could not produce confident family labels for enough of our dataset.

\subsection{Model Selection}
In testing, we found random forests to be accurate classifiers that were easy to train on our data. 
%%While we experimented with other conventional machine learning models like support vector machines and k-nearest neighbors, neither of those models converged to acceptable accuracy and models took many hours to train compared to minutes for the random forests. 
Other work has also found decision tree models to be among the fastest and most accurate among traditional machine learning models for security work \cite{classifiers-compared}\cite{classifier-performance-eval}. 
The particular tree model we found to be most fast and effective was LightGBM\footnote{https://lightgbm.readthedocs.io/en/latest/index.html}, a gradient-boosting tree-based model.

\begin{table}[!t]
\caption{Reduced set of 35 system calls}
\label{reduced-calls}
\begin{center}
\begin{tabular}{lll}
\toprule
read    & write   & creat   \\
open    & openat  & unlink  \\
chdir   & access  & utime   \\
chmod   & ftruncate   & rename  \\
getdents    & fstat   & fstat64 \\
fadvise64   & execve  & rt\_sigaction    \\
rt\_sigprocmask  & kill    &  tgkill  \\
sched\_yield & send    & bind    \\
connect & recvfrom    & poll    \\
epoll\_create    & select  &  ioctl   \\ 
brk & mmap    & mmap2   \\
munmap  & mprotect    & \\
\bottomrule
\end{tabular}
\end{center}
\vspace{-4mm}
\end{table}

\subsection{Feature Processing}
The raw features extracted from the malware experiments were the list of system calls made to the kernel during our execution periods. We decided to perform machine learning analysis using a random forest fed n-grams of system calls from the sequence. This approach has been accurate \cite{ngram-accuracy-study} and showed good and efficient results in our work.

There was some experimentation done to determine a proper feature space. 
%%As of Linux kernel version 3.7, which was current during much of the time our malware samples were collected, there were 393 system calls\cite{syscall-count}. If 3-grams are used as features, this allows 60,235,896 possible features, and 23,491,999,440 if 4-grams are used. These feature spaces are too large to realistically account for every possible combination of system call so our experiments dealt with this in two different ways. 
First, we did not consider every possible combination of calls but only those that were present in all of our experiment runs. Second, to reduce the number of n-grams calculated, we performed the experiments with a limited number of system calls. Limiting the number of calls collected drastically reduces the number of possible n-gram features. Some detail is lost with the discarded calls of course, but careful selection of the retained calls still gives a good view of system activity. Based on \cite{das-fpga-semantics-detection} and observations from our dataset, we determined a short list of system calls that implicate security affecting behavior such as file interaction or network communication. These calls are listed in Table \ref{reduced-calls}.

\begin{figure}[!t]
    \includegraphics[keepaspectratio,width=\columnwidth]{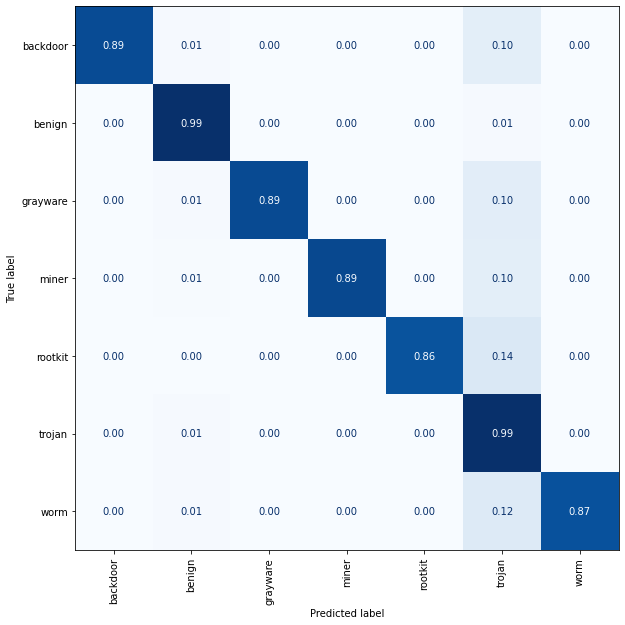}
    \caption{Performance on the baseline experiment set with no background services running}
    \vspace{-2mm}
    \label{no-trojan-baseline-perf}
\end{figure}

While malware injection happens at approximately the same time in each malware experiment, there is a lot of uncertainty in how active the malware is. This has bearing on how n-gram features are extracted. Each malware binary in our final experiment set is confirmed run when it is injected, so features around the midpoint of the execution time will likely contain malware activity. The execution period before malware injection is also certain; the only activity during this period is related to the the baseline services we installed on the execution VMs. As time progresses past malware injection, however, behavior becomes more uncertain. Figure \ref{malware-over-time} shows the phases of the experiments. Features sampled farther and farther from malware injection may or may not actually contain any malware behavior and there is no way to determine this in our feature set.

We experimented with dividing the ten-minute experiment periods into time slices to reduce the amount of time covered by each system call sequence extraction. We tested model performance when all system calls from the 10 minutes were collected together and when the experiment was divided into three, five, and ten slices of equal length. Dividing into ten slices of a minute each was found to give a good balance of model efficacy and extraction performance. Larger slices were less accurate due to the number of non-malware calls increasing and smaller slices were less efficient because feature extraction had to be performed many more times.

We also had to develop a strategy to handle the uncertainty of malware behavior in the latter half of the experiments. The malware is on the system at this time so this time period cannot be considered benign, but the malware may not be active during the entire period and so may not show in the collected features. Labeling that period as a malware class will likely lead to poor classification both during training and classification. We withhold these time slices in training and separate them during classification and reporting.

%\begin{table}[!t]
%\caption{Results for malware classification with n-grams counted across the entire 10-minute execution}
%\begin{center}
%\begin{tabular}{ l l l l l l l }
%\toprule
%Features    & N-gram & Calls   &   Acc. &   Precision   &   Recall    &   F1   \\
%\hline
%Baseline    &   3   &   Complete    &   .92 &   .77 &   .92 &   .83    \\
%Baseline    &   3   &   Reduced     &   .92 &   .73 &   .91 &   .80    \\
%Baseline    &   4   &   Complete    &   .92 &   .76 &   .92 &   .83    \\
%Baseline    &   4   &   Reduced     &   .92 &   .76 &   .93 &   .83    \\
%App         &   3   &   Complete    &   .76 &   .72 &   .75 &   .67    \\
%App         &   3   &   Reduced     &   .85 &   .80 &   .74 &   .76    \\
%\bottomrule
%\end{tabular}
%\end{center}
%\vspace{-2mm}
%\end{table}

\begin{table}[!t]
\caption{Classification performance per-slice for minute-long slices of 3-gram features on baseline data. The bolded row is the only slice in which malware is certainly active}
\label{baseline-performance-per-slice}
\begin{center}
\begin{tabular}{llllll}
\toprule
Minute   &   Label  &   Accuracy    &   Precision   &   Recall  &   F1 \\
1 & Benign & 100.00 & NA & NA & NA \\
2 & Benign & 99.09 & NA & NA & NA \\
3 & Benign & 99.18 & NA & NA & NA \\
4 & Benign & 99.74 & NA & NA & NA \\
5 & Benign & 98.83 & NA & NA & NA \\
6 & \textbf{Malicious} & \textbf{94.30} & \textbf{0.98} & \textbf{0.90} & \textbf{0.95} \\
7 & Malicious & 33.57 & 0.86 & 0.25 & 0.39 \\
8 & Malicious & 55.54 & 0.83 & 0.30 & 0.40 \\
9 & Malicious & 41.82 & 0.84 & 0.28 & 0.41 \\
10 & Malicious & 64.48 & 0.82 & 0.34 & 0.44 \\
\bottomrule
\end{tabular}

\end{center}
\vspace{-4mm}
\end{table}

\section{Results}
\subsection{Result Methodology}
A notable effort was made to determine the suitability of class labels in our dataset. As mentioned in section \ref{classes}, \cite{avclass-raid} discussed the semantic meaning behind certain anti-virus labeling with a focus on the \verb|trojan| label. %The \verb|trojan| label was often used by anti-virus vendors to describe a sample with unspecific behavior when a better classification was impossible to make. In our testing, we noticed a definite overlap in samples between the \verb|trojan| class and other 'similar' classes - mostly the \verb|backdoor| and \verb|virus| classes. 
In Figure \ref{no-trojan-baseline-perf}, it can be observed that several samples belonging to other classes are improperly classified as trojans. Our belief is that this is not due to a lack of model performance, but due to the differences among anti-virus vendors when selecting class labels. If a particular vendor cannot determine a proper class for a piece of malware, it may default to assigning it to the trojan class. Additionally, the described behavior of some of these classes is quite similar. In particular, the classes \verb|trojan|, \verb|rootkit|, and \verb|backdoor| may be reasonably used to describe malware that establishes an illicit backdoor connection to an infected machine. The final class name for samples from these classes is somewhat arbitrarily determined by malware research vendors. 
%\hl{Mis-classification between these classes is not likely to be a major concern because malware behavior in these classes is likely similar. The incident response and remediation plans to remove these malware would likely also be similar.}

\iffalse
\begin{equation}
    \label{precision}
    Precision = \frac{True Positives}{True Positives + False Positives}
\end{equation}

\begin{equation}
    \label{recall}
    Recall = \frac{True Positives}{True Positives + False Negatives}
\end{equation}

\begin{equation}
    \label{f1}
    F1 = 2 * \frac{Precision * Recall}{Precision + Recall}
\end{equation}
\fi
To measure our models' performance, we use four evaluation\footnote{
$Accuracy = \frac{TP + TN}{TP + TN + FP + FN}, Precision = \frac{TP}{TP + FP},\\
Recall = \frac{TP}{TP + FN}, F1-score = 2 \times \frac{Precision \times Recall}{Precision + Recall}$} metrics: accuracy, precision, recall, and F1 score. 
%\hl{While accuracy is sometimes appropriate for data with balanced classes and labels, it can be misleading if the model makes errors on classes with low representation. A model may incorrectly classify all members of a small class and still attain a high accuracy score if the numbers of classes that the model does classify correctly is high enough. Our data is quite imbalanced, so accuracy is inappropriate to measure our performance.} 
Our preferred measure of performance is the model's F1 score. F1 measures the relationship between \textit{precision} and \textit{recall}. Precision measures the number of samples that correctly belong to the predicted class and recall measures how many samples of a class were correctly assigned that class. %Their values are calculated from ratio of true and false positives and negatives according to equations \ref{precision} and \ref{recall}. 
The F1 score is the harmonic mean of precision and recall.

Note that Tables \ref{baseline-performance-per-slice} and \ref{app-performance-per-slice} report NA values for precision, recall, and F1 for time slices that only have one class (benign) present, as having no true positives and false positives or negatives yields zeros in the denominators of the equations. Since we do not label samples in these slices as a positive malicious class, these metrics are difficult to report.

\begin{table}[!t]
\caption{Classification performance per-slice for minute-long slices of 3-gram features on application data. The slices after malware injection have been removed}
\begin{center}
\begin{tabular}{llllll}
\toprule
Minute   &   Label  &   Accuracy    &   Precision   &   Recall  &   F1 \\
1 & Benign & 100.00 & NA & NA & NA \\
2 & Benign & 100.00 & NA & NA & NA \\
3 & Benign & 99.67  & NA & NA & NA \\
4 & Benign & 100.00 & NA & NA & NA \\
5 & Benign & 100.00 & NA & NA & NA \\
6 & Malicious & 74.92  & 0.45 & 0.27 & 0.30 \\
\bottomrule
\end{tabular}
\label{app-performance-per-slice}
\end{center}
\vspace{-4mm}
\end{table}

\begin{table}[!t]
\caption{Detection (binary classification) results for both data sets}
\begin{center}
\begin{tabular}{lllll}
\toprule
Data set    &   Accuracy    &   Precision   &   Recall  &   F1      \\
\hline
Baseline    &   1.00        &   1.00        &   1.00    &   1.00    \\
Application &   98.19       &   .97         &   .97     &   .97     \\
\bottomrule
\end{tabular}
\label{detection-res-table}
\end{center}
\vspace{-4mm}
\end{table}
\subsection{Results Discussed}

Our experiments have determined that this approach is valid for determining the class of a piece of malware on a quiet system, but the approach as implemented is not as effective for classification of malware on a system under heavy load.

The results in Table \ref{baseline-performance-per-slice} shows classification performance on the baseline experiments where the malware is running on a system with no additional configuration after installation. These results indicate that the model has effectively learned to classify these samples based on patterns in their system call usage. There is no malware running on the system until it is injected at the half-way point, so all time slices prior to this point are labeled benign. The time slice where malware is injected and all following slices are labeled as the class to which the malware belongs. While time slices after the inject are labeled as the injected malware class, the malware is not certain to take any action during this period and may not leave any evidence of its presence. The performance metrics for these slices are therefore low. The majority of slices after the inject slice are identified by the model as benign, indicating that the malware was likely not active.

Table \ref{app-performance-per-slice} displays the results of classification for the data collected while a web-server application was active and stress tested on the test machines; see section \ref{exp-env} for experiment details. Clearly the results are much worse. This is attributable to the massive increase in the number of system calls observed on the data sets collected during the baseline and application experiments. The number of features doubles from the baseline to the application dataset. The increase in system calls made by the web service increases the chance that the system calls made by the malware will be dispersed through the benign calls and difficult to pick out.

Despite the inability to accurately classify malware in the application dataset, detection works well. 
Table \ref{detection-res-table} displays the results when all malicious classes are combined into a single \verb|malicious| class. 
While the main purpose of this work is to classify malware on these sequences, the results here do indicate that there is enough feature difference to make conclusions about malware behavior. 
Additional feature processing or model development may accurately classify malware while system call intensive applications are running on the system and is left for further work.

\section{Conclusion and Future Work}
In this work we described a system for attempting to determine the class of malware infecting a system by observing the all system calls made to the kernel. 
This approach is feature efficient because only a single feature feed needs to be collected from the kernel instead of separate feeds from every process.
This work has shown to be effective at determining what class of malware a system is infected with by observing all system calls made to the kernel when other activity on the system is low. 
Future work should continue feature testing and processing to identify ways to improve the performance on systems with high amounts of other activity. 
The detection performance we achieved does indicate that there is enough information in the system call sequences to make some observations on malware activity, so this should be explored further.

\section*{Acknowledgement}
This research is partially supported by NSF Grants 2025682 and 1565562
at TTU, and 2150297 at NCAT.

\bibliographystyle{IEEEtran}
\bibliography{main.bib}

\end{document}